\documentclass[aps,twocolumn]{revtex4}
\usepackage[dvips]{graphicx}
\newcommand{\ri}{{ \rm i }}
\newcommand{\re}{{ \rm e }}
\newcommand{\rd}{{ \rm d }}
\begin{document}               
\title{Wannier-Stark states of a quantum particle in 2D lattices}
\author{M.~Gl\"uck, F.~Keck, A.~R.~Kolovsky\cite{byline} and H.~J.~Korsch}
\affiliation{Fachbereich Physik, Universit\"at Kaiserslautern, D-67653 
         Kaiserslautern, Germany}
\date{\today }

\begin{abstract}
A simple method of calculating the Wannier-Stark resonances in 2D lattices
is suggested. Using this method we calculate the complex Wannier-Stark
spectrum for a non-separable 2D potential realized in optical lattices and
analyze its general structure. The dependence of the lifetime of 
Wannier-Stark states on the direction of the static field (relative
to the crystallographic axis of the lattice) is briefly discussed.
\end{abstract}

\maketitle      

\section{Introduction}

The quantum states of a particle in a periodic potential 
plus homogeneous field (known nowadays as the Wannier-Stark states, 
WS-states in what follows) are
one of the long-standing problems of single-particle quantum mechanics. 
The beginning of the study of this problem dates back to the paper by 
Bloch of 1929, followed by contributions of Zener, 
Landau, Wannier, Zak and many others \cite{history}.
In the late eighties the problem got a new impact by the invention of semiconductor
superlattices. The unambiguous observation of the WS-spectrum in
a semiconductor superlattice \cite{mendez} ended a long theoretical
debate about the nature of WS-states, and now it is commonly accepted
that they are the {\em resonance states} of the system. Besides, WS-states were recently studied in a system of cold atoms in 
an optical lattice \cite{atoms} and some other (quasi) one-dimensional
systems.

Although WS-states are resonances, i.e.~metastable states, in
the theoretical analysis of related problems they were usually approximated
by stationary states (one-band, tight-binding, and similar approximations).
Beyond the one-band approximation, WS-states in the semiconductor and optical 
lattices were studied in recent papers \cite{preprint} and \cite{papers} by using the scattering matrix approach of Ref.~\cite{PRL1}
(see also Ref.~\cite{JOB2} for details). This approach actually solves  
the one-dimensional Wannier-Stark problem and supplies exhaustive 
information about 1D WS-states. In the present letter we extend the 
method of Ref.~\cite{PRL1,JOB2} to the case of {\em two-dimensional}
lattices. For the first time we find the complex spectrum of 2D WS-states 
and analyze its general structure.

\begin{figure}
\center
\includegraphics[width=8cm]{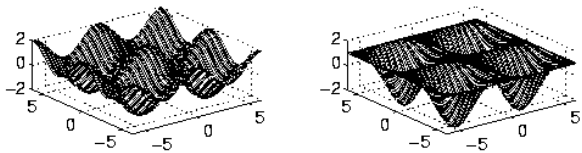}
\caption{Potential energy (\ref{2}) for $\epsilon=0$ (a) and
$\epsilon=1$ (b).}
\label{lfig1}
\end{figure}To be concrete, we choose the following system:
\begin{equation}
\label{1}
H={\bf p}^2/2+V({\bf r})+{\bf F}\cdot{\bf r} \;, \qquad {\bf r}=(x,y) \;,
\end{equation}
\begin{equation}
\label{2}
V({\bf r})=\cos x+\cos y-\epsilon\cos x\cos y \;,
\end{equation}
where $0\le\epsilon\le1$ \cite{hbar}. 
Two limiting cases $\epsilon=0$ and $\epsilon=1$ 
correspond to an `egg crate' potential, for which the system is separable,
and a `quantum well' potential, where the coupling between two degrees of
freedom is maximal (see Fig.~\ref{lfig1}). Let us also note that the choice 
$\epsilon=1$ corresponds to a 2D optical potential created by
two standing laser waves crossing at right angle. 
Thus the results presented below can be directly applied to the system
of cold atoms in a 2D optical lattice.

\section{2D Wannier-Bloch spectrum}

We briefly recall the key points of the 1D theory. 
The spectrum of the Bloch particle in the presence of a static field
consists of several sets of equidistant levels 
\begin{equation}
{\cal E}_{\alpha,l}=E_\alpha+2\pi Fl-\ri \Gamma_\alpha/2 \;,
\label{18}
\end{equation}
known as Wannier-Stark ladders of resonances. In Eq.~(\ref{18}), $2\pi$
stands for the lattice period, $F$ is the amplitude of the static force, $l=0, \pm 1, \dots$ is 
the site index and the index $\alpha=0,1,\ldots$ labels different ladders. 
The lifetime of WS-states $\Psi_{\alpha,l}(x)$ 
is defined by the resonance width 
$\Gamma_\alpha$ as $\tau_\alpha=\hbar/\Gamma_\alpha$. Typically, the
lifetime $\tau_\alpha$ rapidly decreases with increasing index $\alpha$.
Because of this only the first few WS-ladders are of physical importance.

Along with the WS-states $\Psi_{\alpha,l}(x)$, one can also introduce Wannier-Bloch states (WB-states) by
\begin{equation}
\psi_{\alpha,k}(x)=\sum_l \Psi_{\alpha,l}(x) \exp(\ri 2 \pi k l) \;.
\label{19}
\end{equation}
As follows from the definition (\ref{19}), the continuous evolution of
WB-states obeys the equation  $\psi_{\alpha,k}(x,t)=
\exp(-\ri{\cal E}_ \alpha t/\hbar)\psi_{\alpha,k-Ft/\hbar}(x)$,
where ${\cal E}_\alpha=E_\alpha-\ri\Gamma_\alpha/2$.
Thus, WB-states can be alternatively defined as the eigenfunction of the 
evolution operator over the Bloch period $T_B=\hbar/F$ \cite{niu2}. (Note that the
eigenvalues of the evolution operator form degenerate bands
${\cal E}_\alpha(k)={\cal E}_\alpha$).
Additionally, to ensure that $\psi_{\alpha,k}(x)$ are resonance
states of the system, the eigenvalue equation for the evolution operator
should be accomplished by the specific non-hermitian boundary condition.
It was proven in Ref.~\cite{JOB2} that the required boundary conditions
are imposed by the truncation of the evolution operator matrix in the
momentum representation.

We proceed with the two-dimensional case. As mentioned above, WB-states in a 1D lattice
can be defined as the non-hermitian eigenstates of the evolution 
operator over one Bloch period. 
In the 2D problem there are
two different Bloch periods associated with the two components of the
static field. 
Therefore the notion of the WB-states can be introduced
only in the case of commensurate periods, i.e., in the case
of `rational' direction of the field ($q, r$ are coprime integers):
\begin{equation}
\label{21}
F_x=\frac{qF}{(r^2+q^2)^{1/2}} \;,\quad F_y=\frac{rF}{(r^2+q^2)^{1/2}} \;.
\end{equation}
Provided condition (\ref{21}) is satisfied, we define 2D WB-states
as the non-hermitian eigenfunctions of the system evolution operator
over the common Bloch period $T_B=(r^2+q^2)^{1/2}\hbar/F$.
Using the Kramers-Henneberger transformation, which is just the gauge which transforms the static term into the vector potential, the 
evolution operator can be presented in the form
\begin{equation}
\label{23}
\widehat{U}(T_B)=\re^{-\ri qx}\,\re^{-\ri ry} \, \widehat{\exp} \left( -\frac{\ri}{\hbar} \int_0^{T_B} \rd t \, \tilde{H}(t) \right) \, ,
\end{equation}
\begin{equation}
\tilde{H}(t)= \frac{(\hat{p}_x-F_xt)^2}{2}+\frac{(\hat{p}_y-F_yt)^2}{2}+V(x,y) \, , 
\end{equation}
which reveals its translational invariance (the hat over the exponent sign denotes time ordering).
Alternatively, we can rotate the coordinates so that the direction of the field coincides with the $x'$-axis:
\begin{equation}
\label{24}
x'=\frac{qx+ry}{(r^2+q^2)^{1/2}} \;,\quad 
y'=\frac{qy-rx}{(r^2+q^2)^{1/2}}\;.
\end{equation}
Transformation (\ref{24}) introduces a new lattice period 
$a=2\pi(r^2+q^2)^{1/2}$ and reduces the size of the original
Brillouin zone $s=r^2+q^2$ times. 
Associated with the new lattice period is a new Bloch time 
$T_a = (r^2 + q^2)^{-1/2} \, \hbar / F$, 
which is $s$ times shorter than the original Bloch time $T_B$.
Using $\hat{p}'_x=-\ri \hbar\partial/\partial x'$ and
$\hat{p}'_y=-\ri \hbar\partial/\partial y'$, the time evolution operator
over the new Bloch time $T_a$ in the rotated coordinates has the form
\begin{equation}
\label{25}
\widehat{U}'(T_a)=\re^{-\ri 2 \pi x'/a} \, \widehat{\exp} \left( -\frac{\ri}{\hbar} \int_0^{T_a} \rd t \, \tilde{H}'(t) \right) \, ,
\end{equation}
\begin{equation}
\tilde{H}'(t)=\frac{(\hat{p}_x' -Ft)^2}{2} + \frac{\hat{p}_y'}{2} + V(x', y') \, .
\end{equation}
%
Then, presenting the wave function as 
\begin{equation}
\label{26}
\psi({\bf r'})=\re^{i{\bf k'r'}}
\sum_{\bf n'} c_{\bf n'}\langle{\bf r'}|{\bf n'}\rangle 
\;,\quad
\langle{\bf r'}|{\bf n'}\rangle=\frac{1}{a} \, 
\re^{\ri 2\pi {\bf n' \cdot r'}/a} \;,
\end{equation}
we get the matrix equation
\begin{equation}
\label{27}
\sum_{\bf m'} U'^{({\bf k'})}_{{\bf n'}{\bf m'}} c_{\bf m'}
=\re^{-\ri {\cal E} T_a / \hbar}  c_{\bf n'} \;,
\end{equation}
where $U'^{({\bf k'})}_{{\bf n'}{\bf m'}}$ denotes the $k'$-dependent matrix 
elements of the operator (\ref{25}):
\begin{equation}
\label{28}
U'^{({\bf k'})}_{{\bf n'}{\bf m'}} =
\langle{\bf n'}| \re^{-\ri{\bf k' \cdot r'}} \, \widehat{U}'(T_a) \, 
\re^{\ri {\bf k' \cdot r'}}|{\bf m'}\rangle \;.
\end{equation}
Similar to the 1D case, the truncation of the infinite unitary matrix (\ref{28}), 
\begin{equation}
\label{29}
|n'_x|,|m'_x|\le N \rightarrow\infty \;,
\quad |n'_y|,|m'_y|\le M \rightarrow\infty \;,
\end{equation}
which is presumed in the numerical calculations, automatically imposes 
the non-hermitian boundary condition along the $x'$-direction.
(Truncation of the matrix over the index $n'_y$, $m'_y$ does not 
change the hermitian 
boundary condition along the $y'$-direction.) Then the eigenvalues ${\cal E}$ 
obtained by numerical diagonalization of the truncated matrix correspond to 
the quantum resonances.

In the transformed coordinates, the unit cell with area $a^2 = (2\pi)^2 s$ 
contains $s$ different sublattices, 
and each of them supports its own WB-states. The sublattices are
related by primitive translations of the unrotated lattice, and
correspondingly the energies of their WB-states differ by multiples
of $aF/s$. 
Furthermore, as function of the quasimomentum, the energies
${\cal E}={\cal E}_\beta^{(i)}(k'_x,k'_y)$ (here $\beta=0,1,\ldots$ is the 
`Bloch band' index and $i=1,\ldots,s$ is the sublattice index) do not depend
on $k'_x$. This follows from the fact that a change of $k'_x$ in 
Eq.~(\ref{28}) can be compensated by shifting the time origin in Eq.~(\ref{25}).
For the $y'$-degree of freedom the Bloch theorem can be applied,
and therefore ${\cal E}_\beta^{(i)}(k'_x,k'_y)$ is a periodic 
function of $k'_y$ with generally nonzero amplitude $\Delta{\cal E}_\beta$. 
Thus, assuming a rational direction of the field, in each fundamental 
energy interval $a F$, the static field induces $s=r^2+q^2$ identical 
sub-bands, separated by the energy interval $a F/ s$. Simultaneously, 
the size of the Brillouin zone is reduced by a factor $s$.
This result resembles the one obtained for a 
1D lattice affected by a time-periodic perturbation \cite{niu} or that for a 2D lattice in a magnetic field \cite{azbel}. In
these cases -- provided the condition of comensurability between 
the Bloch period and the period of the driving force or the condition of `rationality' for the magnetic flux through a unit cell, respectively, is fulfilled -- the (quasi)energy 
spectrum of the system has a similar structure.

We conclude this section with a remark concerning the numerical procedure.
Although the reduced Brillouin zone approach described above is the most
consistent, we found it more convenient to diagonalize the evolution operator
without preliminary rotation of the coordinate. In other words, in order to find the
WB-spectrum, we solve the eigenvalue equation (\ref{27}) with the
truncated matrix constructed on the basis of the operator (\ref{23}).
As a result of the diagonalization, one obtains 
eigenvalues ${\cal E}_\beta (k_x,k_y)$ with quasimomentum
${\bf k}=(k_x,k_y)$ defined in the original Brillouin zone. Because the
WB-bands are uniform along the direction of the field,
${\cal E}_\beta (k_x,k_y)$ is a periodic function of both
$k_x$ and $k_y$ with periods $1/r$ and $1/q$ respectively. 
The energies obtained in this way can then be  
used to construct the complete WB-spectrum 
${\cal E}_\beta^{(i)}(k'_x,k'_y)$, $i=1,\ldots,s$. In the next
section we present results of a numerical calculation of the
dispersion relation ${\cal E}_\beta (k_x,k_y)$ for the periodic potential
(\ref{2}) and moderate values of the static field ${\bf F}=(F_x,F_y)$,
$|{\bf F}|=F=const$.


\section{Numerical results}

It is instructive to begin with the separable case $\epsilon=0$. 
In this case, 2D WB-states are given by the product of 1D states
and 2D WB-energies are just the sum of 1D energies. In what follows we 
restrict ourselves to analyzing only the ground band.
First we consider the real part of the spectrum $E_0={\rm Re}({\cal E}_0)$.
\begin{figure}
\center
\includegraphics[width=8cm]{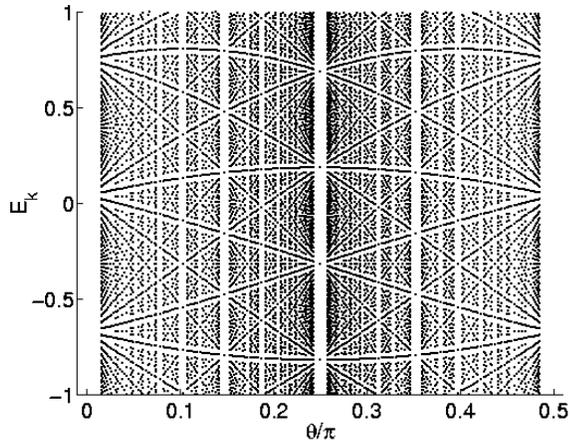}
\caption{Position of the ground WB-band repeated by the subband energy
interval $2\pi F(r^2+q^2)^{-1/2}$ as a function of the field direction $\theta=\arctan(r/q)$ (parameters $\hbar=2$, $F=0.08\sqrt{2}$, $\epsilon=0$, integers $q, r\le 21$).}
\label{lfig3}
\end{figure} 

It was shown in the previous section that for rational directions of the field the 
ground WB-subbands repeat with energy splitting $aF/s$. As an example, Fig.~\ref{lfig3} shows
the relative positions of these subbands as a function of the angle 
$\theta=\arctan(r/q)$ for $\hbar=2$ and $F=0.08\sqrt{2}$. We recall that
in the considered case of a separable potential the bands have zero
width for any $\theta \neq 0, \pi/2$.

\begin{figure}
\center
\includegraphics[width=8cm]{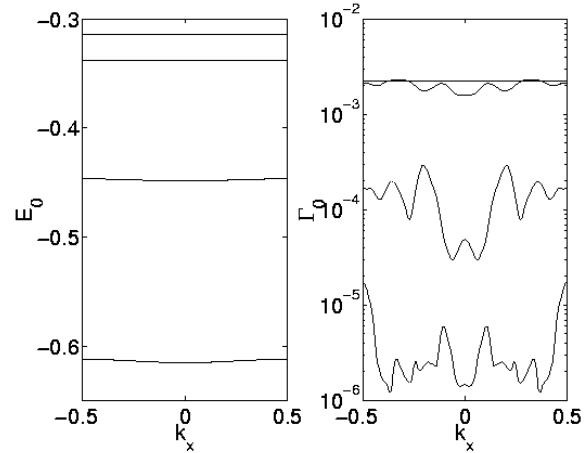}
\caption{Real (left) and imaginary (right) parts of the dispersion relation
${\cal E}_\beta(k_x,k_y)$ for the ground WB-states
and different values of the potential parameter $\epsilon=0$, $0.1$,$0.5$, and $1$ (from top to bottom). 
The system para\-meters are $\hbar=2$, $F_x=F_y=0.08$, and $k_y=0$.}
\label{lfig4}
\end{figure}
The main difference between  separable and non-separable potentials
is that the subbands $E^{(i)}_0({\bf k})$ have a finite width in the
latter case. This is illustrated by Fig.~\ref{lfig4}(a) which shows the
dispersion relation $E_0(k_x,k_y=0)$  
for the potential (\ref{2}) with (from top to bottom) $\epsilon=0$, $0.1$, 
$0.5$, and $1$. The direction of the field is $\theta=\pi/4$, 
i.e.~$r=q=1$. The amplitude of the static field and the value 
of the scaled Planck constant are the same as in Fig.~\ref{lfig3}.
It is seen in Fig.~\ref{lfig4}(a) that the WB-bands gain a finite width as $\epsilon$ 
is increased. We also calculated the dispersion relation $E_0(k_x,k_y=0)$ for
different angles $\theta=\arctan(r/q)$, with $r,q\le6$. It was found that the band 
widths $\Delta E_0=\Delta E_0(r,q)$ are typically much smaller than the mean
energy separation between the subbands. Thus, for practical purpose, one
can neglect the band width for the real part of the spectrum. (An exception 
is the case $\theta=0,\pi/2$ where the width of the WB-bands approximately coincides
with the width of the Bloch band in the absence of the static field.)
Neglecting the width of the bands they were found to form a structure
similar to that shown in Fig.~\ref{lfig3}.

\begin{figure}
\center
\includegraphics[width=8cm]{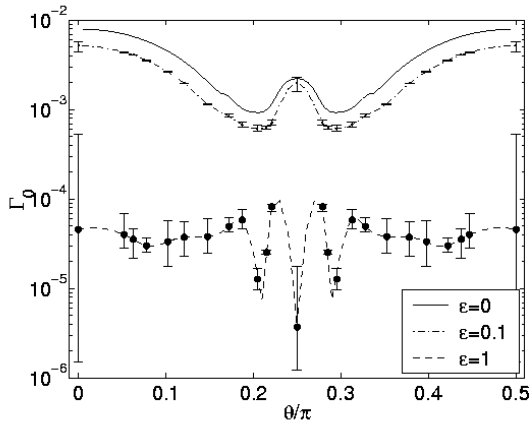}
\caption{Decay rate of the ground WB-states as a function 
of the field direction $\theta$ in the case of separable potential ($\epsilon=0$, solid curve).
The dashed and dashed-dotted lines are an interpolation to arbitrary $\theta$ 
of the mean decay rate calculated for some rational directions of the field 
(dots) for $\epsilon=0.1$ and $\epsilon=1$, respectively. The maximum and minimum values 
of the decay rate for these angles are indicated by the `error' bars.}
\label{lfig5}
\end{figure}

We proceed with the analysis of the decay rate of the WB-states, which is determined
by the imaginary part of the complex energy, $\Gamma_0=-2{\rm Im}({\cal E}_0)$.
In the case of a separable potential the dependence $\Gamma_0=\Gamma_0(F, \theta)$ 
is obviously given by the equation
\begin{equation}
\label{31}
\Gamma_0(F, \theta)=\Gamma'_0(F\cos\theta)+\Gamma'_0(F\sin\theta) \;,
\end{equation}
where $\Gamma'_0(F')$ stands for the width of 1D WS-resonances. For the
parameters used ($\hbar=2$ and $F=0.08\sqrt{2}$) the dependence
(\ref{31}) is shown in Fig.~\ref{lfig5} by a solid line. The maximum around 
$\theta=\pi/2$ originates from a peak-like behavior of $\Gamma'_0(F')$
and is explained by the phenomenon of 1D resonant tunneling \cite{JOB2}. 

For a non-separable potential and rational direction of the field
the decay rate depends on the quasimomentum. For the particular case
$\theta=\pi/4$ this dependence is depicted in Fig.~\ref{lfig4}(b). 
We would like to note the complicated behavior of $\Gamma_0({\bf k})$.
The oscillating character of the decay rate is an open problem for the 
present day. Because the decay rate depends on the quasimomentum it
might be convenient to introduce the notion of $\bar{\Gamma}_0$, 
where the average is taken over the reduced Brillouin zone. 
The dots in Fig.~\ref{lfig5} show the values of $\bar{\Gamma}_0$ for 
some rational direction of the field and two different values of $\epsilon$. 
It is seen that for a small $\epsilon=0.1$ the ratio 
$\Delta \Gamma_0/\bar{\Gamma}_0$ is small and the obtained dependence
$\bar{\Gamma}_0=\bar{\Gamma}_0(r,q)$  essentially reproduces that
of the separable case. However, this is not valid for $\epsilon=1$,
where the decay rate varies wildly. Thus, in the case of strong coupling 
between two degrees of freedom the description of WS-state by a
mean decay rate is insufficient.

\section{Conclusion}

We studied Wannier resonances in a 2D system, mainly discussing the
complex energy spectrum of the Wannier-Bloch states. However, because the latter
are related to the Wannier-Stark states by a Fourier transformation,
the obtained results can be easily reformulated in terms of the
Wannier-Stark resonances. Then the following is valid. (i) Neglecting
the asymptotic tail, WS-states are localized functions along the direction
of the field. (This follows from the degeneracy of WB-bands along the field
direction.) (ii) For any rational direction of the field [see Eq.~(\ref{21})]
WS-states are Bloch waves in the transverse direction. (iii) For
a non-separable potential the corresponding energy bands have a finite width.
(iv) For the {\em real} part of the spectrum, the band widths are small and can be
well neglected for $r,q>1$.

We also found a nontrivial dependence of the resonance width (inverse
lifetime of WS-states) on the direction of the field. Because the value
of the resonance width defines the decay of the probability, a complicated
behavior of the survival probability is expected when the direction of
the field is varied. The detailed study of the probability dynamics
is reserved for future publication.

%


\begin{references}
\bibitem{hbar}
Through the paper we use dimensionless variables where the amplitude
of the static field and the scaled Planck constant are the independent 
parameters of the system. 
Alternatively, one can set $\hbar=1$ and introduce
the notion of the scaled amplitude $V_0\ne1$ for the potential (\ref{2}).
\end{references}
\end{document}